# Influence of gallium content on $Ga^{3+}$ position and photo- and thermally stimulated luminescence in $Ce^{3+}$ - doped multicomponent $(Y,Lu,)_3Ga_xAl_{5-x}O_{12}$ garnets

V. Babin[1], M. Buryi[1], K. Kamada[2], V.V. Laguta[1,3], M. Nikl[1], J. Pejchal[1], H. Štěpánková[4], A. Yoshikawa[5], Y. Fomichov[3,4], Yu. Zagorodniy[3,4], S. Zazubovich[6]

[1]*Institute of Physics AS CR, Cukrovarnicka 10, 16253 Prague 6, Czech Republic*
[2]*New Industry Creation Hatchery Center, Tohoku University, 2-1-1 Katahira, Aoba-ku, Sendai 980-8577, Japan*
[3]*Institute for Problems of Materials Science NAS Ukraine, Krjijanovsky 3, 03142 Kyiv, Ukraine*
[4]*Charles University in Prague, Faculty of Mathematics and Physics, V Holešovičkach 2, 180 00 Prague 8, Czech Republic*
[5]*Institute for Materials Research, Tohoku University, 2-1-1, Katahira, Aoba-ku, Sendai, Miyagi 980-8577, Japan*
[6]*Institute of Physics, University of Tartu, W. Ostwaldi 1, 50411 Tartu, Estonia*

**Abstract**

Photoluminescence, thermally stimulated luminescence (TSL) and EPR characteristics of the $Ce^{3+}$ - doped single crystals of multicomponent $Y_1Lu_2Ga_xAl_{5-x}O_{12}$ and $Lu_3Ga_xAl_{5-x}O_{12}$ garnets with different Ga contents (x = 0, 1, 2, 3, 4, 5) excited in the $Ce^{3+}$ - related absorption bands are investigated in the 9 - 500 K temperature range. The distribution of $Ga^{3+}$ and $Al^{3+}$ ions in the crystal lattice is determined by the NMR method. The relative number of $Ga^{3+}$ ions in the tetrahedral crystal lattice sites, the maxima positions of the TSL glow curve peaks and the corresponding trap depths are found to decrease *linearly* with the increasing Ga content. At the same time, the reduction of the activation energy $E_a$ of the TSL glow curve peaks creation under irradiation in the $4f - 5d_1$ absorption band of $Ce^{3+}$ is strongly *nonlinear*. To explain this effect, the suggestion is made that $E_a$ is the energy distance between the excited $5d_1$ level of $Ce^{3+}$ and a defect level located between the $5d_1$ level and the bottom of the conduction band and arising from the $Ga^{3+}$ ion perturbed by the nearest neighboring $Ce^{3+}$ ion. The electrons thermally released from the excited $Ce^{3+}$ ions are suggested to be trapped at the perturbed $Ga^{3+}$ ions resulting in the appearance of electron $Ga^{2+}$ centers. In spite of the fact that the paramagnetic $Ga^{2+}$ ions were not detected by EPR, the described above process was found for $Fe^{3+}$ impurity ions, namely the electron transfer from the $5d_1$ excited levels of $Ce^{3+}$ to $Fe^{3+}$ was directly detected by EPR.

# 1. Introduction

Multicomponent garnets of general chemical formula $(Y,Lu,Gd)_3(Ga,Al)_5O_{12}$:Ce are relatively new ultraefficient complex oxide scintillators with extremely high light yield extending 50 000 ph/MeV (see, e.g., review [1]). These scintillation materials have been discovered in 2011 [2,3], and a huge number of papers appeared in the recent years devoted to the study of the photo- and thermally stimulated luminescence and scintillation characteristics of the single crystals and epitaxial films of the $Ce^{3+}$ - doped multicomponent garnets of different composition (see, e.g., [1-31] and references therein).

The preparation of multicomponent solid solutions became of great interest due to the possibility to enhance their scintillation performance by the band gap and defect engineering [1-4]. Such engineering of the electronic band structure of the host and positioning of the $Ce^{3+}$ levels in its forbidden gap results in the reduction of a negative influence of electron traps on the energy transfer processes as well as in the enhancement of the scintillation light yield as compared to the single-component compositions (see, e.g., [1-6,8,9,14-16,20,22,23,32] and references therein).

Under irradiation of the $Ce^{3+}$ - doped multicomponent garnets at appropriate temperatures in the 4f - $5d_1$ (around 2.8 eV) or 4f – $5d_2$ (around 3.6 eV) absorption bands of $Ce^{3+}$, the electrons can be thermally released from the $5d_1$ or $5d_2$ excited levels of $Ce^{3+}$, be trapped at different traps, thermally released from these traps, and finally recombine with the optically created $Ce^{4+}$ hole centers. As a result, the $Ce^{3+}$ - related TSL appears. The activation energy $E_a$ for a TSL glow curve peak creation can be determined from the dependence of the TSL peak intensity ($I_{TSL}$) on the irradiation temperature ($T_{irr}$). In case the considered TSL peak appears as a result of the electron release from the $5d_1$ or $5d_2$ level of $Ce^{3+}$ into the conduction band (CB), the $E_a$ value for the TSL peak creation should correspond to the energy distance between the $5d_1$ or $5d_2$ excited level of $Ce^{3+}$ and the bottom of the CB. In this case, the $E_a$ value should decrease with the increasing Ga content in correspondence with the decreasing CB bottom energy.

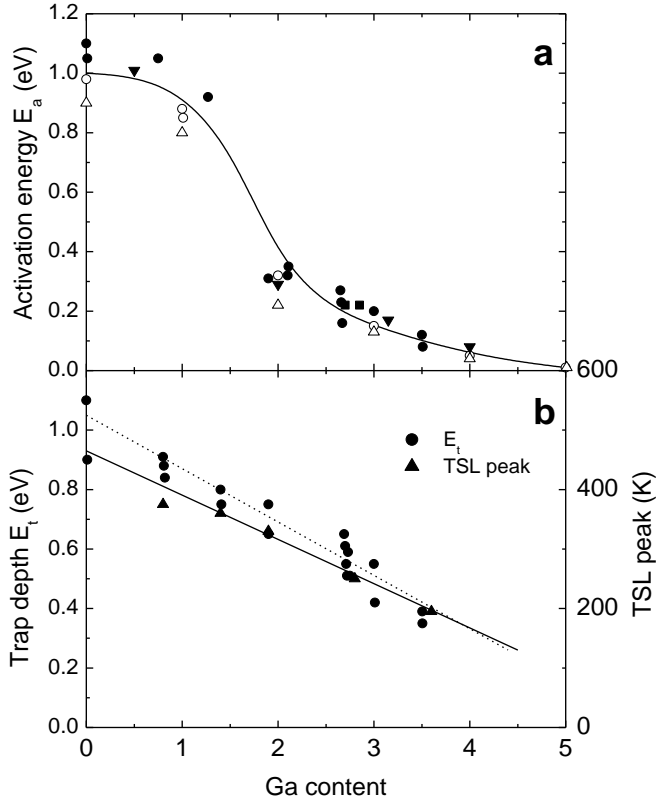

**Fig. 1.** Dependences of (a) the activation energy $E_a$ of the TSL peaks creation and (b) the position of the high-temperature TSL peak (filled triangles) and the corresponding trap depth $E_t$ (filled circles) on the Ga content. The data in Fig. 1a are obtained for the $Lu_2Y_1(Ga,Al)_5O_{12}$:Ce (empty circles), $Gd_3(Ga,Al)_5O_{12}$:Ce (filled triangles) and $Lu_3(Ga,Al)_5O_{12}$:Ce (empty triangles) single crystals, for the $Gd_3Ga_xAl_{5-x}O_{12}$:Ce and $Gd_3Ga_xAl_{5-x}O_{12}$:Ce,Mg single crystals with x = 2.831 and x = 2.694, respectively (filled squares) studied in [30], and for the $(Gd,Lu)_3(Ga,Al)_5O_{12}$:Ce epitaxial flms (filled circles) studied in [25]. The data presented in Fig. 1b are obtained for the epitaxial films of $(Gd,Lu)_3(Ga,Al)_5O_{12}$:Ce studied in [25].

The $E_a$ values obtained by this way for the epitaxial films of $Ce^{3+}$ - doped multicomponent $(Lu,Gd)_3Ga_xAl_{5-x}O_{12}$ garnets of different composition studied in [25] are presented in Fig. 1a (filled circles). It is evident, that the $E_a$ dependence on the Ga content is
strongly nonlinear. The strongest reduction of the $E_a$ value takes place between x = 1.4 and x = 2. If we assume that the $E_a$ value is always equal to the energy distance between the $5d_1$ or $5d_2$ level and the bottom of the CB and the reduction of the $E_a$ value is caused by the $Ga^{3+}$ - induced lowering of the CB bottom energy [1-4], the same nonlinear dependence on the Ga content should be observed also for the TSL peaks positions as well as for the corresponding trap depths $E_t$. However, the data presented in Fig. 1b indicate that this is not so at least in case of the $(Lu,Gd)_3Ga_xAl_{5-x}O_{12}$:Ce epitaxial films studied in [25].

To understand the reason of this strong disagreement, the $Y_1Lu_2Ga_xAl_{5-x}O_{12}$:Ce and $Lu_3Ga_xAl_{5-x}O_{12}$:Ce single crystals with different Ga contents (x) are studied in this work by the photo- and thermally stimulated luminescence methods. Some data obtained for the $Gd_3Ga_xAl_{5-x}O_{12}$:Ce single crystals in [33] are taken into account as well. The obtained results are compared with the data reported for the $(Lu,Gd)_3Ga_xAl_{5-x}O_{12}$:Ce epitaxial films in [25]. To clarify the dependence of the position of $Ga^{3+}$ ions (in tetrahedral or octahedral Al sites) on the Ga content, the position of $Ga^{3+}$ ions in the crystal lattice of $Lu_3Ga_xAl_{5-x}O_{12}$:Ce is investigated by the NMR method. The possibility of formation of electron centers under irradiation in the $Ce^{3+}$ - related absorption bands is investigated by EPR.

## 2. Experimental procedure

The single crystals of $Y_1Lu_2Ga_xAl_{5-x}O_{12}$:Ce with the Ce content of 1 at.% and $Lu_3Ga_xAl_{5-x}O_{12}$:Ce with the Ce content of 0.2 at.% investigated in this paper were prepared by the micro-pulling down method in Sendai and Prague, respectively. Some data were also obtained for the $Gd_3Ga_xAl_{5-x}O_{12}$:Ce single crystals with the Ce content of 0.2 at.% prepared by the micro-pulling down method in Sendai. In more detail, the luminescence characteristics of the latter crystals are investigated in a separate work [33]. The Ga content in the crystals studied was x = 0, 1, 2, 3, 4, 5.

The steady-state emission and excitation spectra in the 85 - 500 K temperature range were measured using a setup, consisting of the LOT - ORIEL xenon lamp (150 W), two monochromators (SF - 4 and SPM - 1) and a nitrogen cryostat. The luminescence was detected by a photomultiplier (FEU - 39 or FEU - 79) connected with an amplifier and recorder.

Thermally stimulated luminescence glow curves $I_{TSL}(T)$ were measured at the same setup with a heating rate of 0.2 K/s in the 85 - 500 K temperature range after selective UV irradiation of the crystals at different temperatures $T_{irr}$ with different irradiation photon energies $E_{irr}$. A crystal located in a nitrogen cryostat was irradiated with the LOT - ORIEL xenon lamp (150 W) through a monochromator SF-4. The TSL glow curves were measured with the monochromator SPM - 1 and detected with the photomultiplier FEU - 39 and recorder. For each TSL glow curve peak, the TSL peak creation spectrum, i.e., the dependence of the maximum TSL intensity ($I_{TSL}^{max}$) on the irradiation photon energy $E_{irr}$, was measured. From the dependence of the maximum TSL intensity ($I_{TSL}^{max}$) on the irradiation temperature $T_{irr}$, the activation energy $E_a$ for the TSL peak creation was determined. To determine the trap depth $E_t$ corresponding to each TSL peak, the partial cleaning method was used (for more details, see, e.g., [34] and references therein). The crystal, irradiated at the temperature $T_{irr}$, was cooled down to 85 K, heated up to a temperature $T_{stop}$, then quickly cooled down to 85 K and the TSL glow curve was recorded. In the next cycle, the same procedure was repeated for the different temperature $T_{stop}$, etc. From the slope of the $\ln(I_{TSL})$ as a function of the reverse temperature (1/T), the $E_t$ value was calculated.

NMR measurements were carried out at room temperature using a commercial Bruker Avance III HD solid-state NMR spectrometer equipped with high speed (magic-angle spinning) MAS probe at the Larmor frequency 130.40 MHz and 152.62 MHz for the $^{27}$Al and $^{71}$Ga isotopes, respectively. $^{27}$Al MAS NMR spectra were measured by the single pulse sequence radio-frequency (RF) field strength of 110 KHz. Short pulses of 0.38 µsec, corresponding to the rotation of the net magnetization on the angle of π/12 were used to ensure the correctness of quantitative measurements of aluminum concentrations from different lattice positions with different values of the quadrupole constants. From 512 up to 3072 scans with recycle delay of 80 s were accumulated for each sample depending on the Al concentration. Spectral width of 312 KHz with spinning speed of 22 KHz was used to observe an undistorted central transition of $^{27}$Al in both the tetrahedral and octahedral positions. $^{27}$Al NMR spectra are referenced to external standard $Al(NO_3)_3$.

Since the $^{71}$Ga NMR spectrum is much broader than available presently spinning rates in MAS experiment, it can not be narrowed by MAS technique. Therefore, to distinguish two occupation sites of Ga ions, single-crystal NMR spectra were measured by utilizing the 90x − τ − 90y − τ spin echo pulse sequence. Four phase (*xx*, *xy*, *x-x*, *x-y*) 'exorcycle' phase cycling was used to form echoes with minimal distortions due to anti-echoes, ill-refocused signals and piezo-resonances. In the case of broad $^{71}$Ga NMR spectrum (up to 300 kHz for the 1/2 ↔ -1/2 central transition), only selective excitation takes place, but to exclude the influence of the angle dependence of quadrupolar interactions on the signal intensity, π/8 pulse lengths of 0.5 µs were used with 5 s delays between scans.

EPR spectra were measured in the X-band (9.4 GHz) with a commercial Bruker X/Q-band E580 FT/CW ELEXSYS spectrometer within the 10–297 K temperature range using an Oxford Instruments ESR900 continuous flow cryostat. A Cd/He laser working at 325 and 442 nm was used for the optical irradiation of the samples directly in the cavity of the spectrometer.

3. **Experimental results and discussion**

   *3.1. NMR investigation of Ga and Al distribution in $Lu_3Ga_xAl_{5-x}O_{12}$:Ce*

   In accordance with the previous studies [29,35,36], $^{27}$Al MAS NMR allows easy separation of spectral lines from Al ions in the tetrahedral and octahedral positions of garnet structure. $^{27}$Al

MAS NMR spectra of $Lu_3Ga_xAl_{5-x}O_{12}$:Ce single crystals obtained with the spinning rate of 22 KHz are presented in Fig. 2. This rotation speed exceeds the line width of the static spectrum, therefore all magnetization from the central transition is presented in only one line. Two well separated spectral lines from Al belonging to the tetrahedral (Al(IV)) and octahedral (Al(VI)) positions are well visible for all Ga concentrations. The Al(IV)/Al(VI) occupation ratio can be calculated from the areas under corresponding spectral lines.

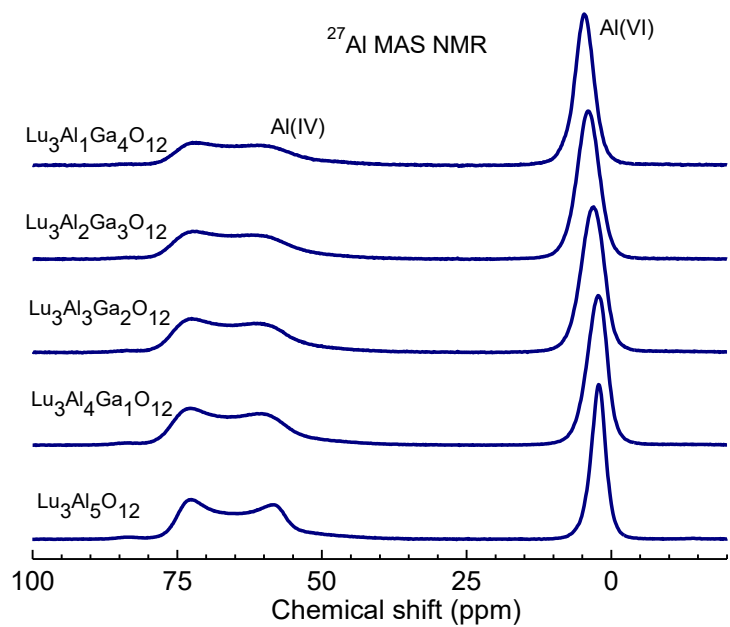

**Fig. 2**. $^{27}$Al MAS NMR spectra of $Lu_3Ga_xAl_{5-x}O_{12}$:Ce compounds obtained at rotation speed of 22 KHz. Only frequency region of the 1/2 ↔ -1/2 central transition is shown.

In order to obtain correct values of the Al(IV)/Al(VI) ratio one should subtract part of the magnetization belonging to the satellite transitions from the central transition region. It was done by simulating of the experimentally accumulated spectrum in the broad frequency range of approximately ±1 MHz to account all $^{27}$Al transitions. Actual ratios of Al(IV)/Al(VI) in $Lu_3Ga_xAl_{5-x}O_{12}$:Ce compounds are presented in Table 1. These ratios are considerably smaller in the mixed crystals than that of 3/2 expected for random distribution of Al at tetrahedral and octahedral sites. The difference from the 3/2 value increases with the Ga concentration increase indicating the preference of tetrahedral sites for Ga.

**Table 1**. Occupation ratios Al(IV)/Al(VI) and Ga(IV)/Ga(VI) in $Lu_3Ga_xAl_{5-x}O_{12}$:Ce solid solutions determined from $^{27}$Al and $^{71}$Ga NMR spectra.

| Compound | Al(IV)/Al(VI) | Ga(IV)/Ga(VI) | Ga(IV)/Ga(VI)[*)] |
|---|---|---|---|
| LuAl$_5$O$_{12}$ | 1.51 | | |
| LuAl$_4$Ga$_1$O$_{12}$ | 1.19 | 3.6 | 4.8 |
| LuAl$_3$Ga$_2$O$_{12}$ | 1.01 | 3.0 | 2.9 |
| LuAl$_2$Ga$_3$O$_{12}$ | 0.84 | 2.15 | 2.29 |
| LuAlGa$_4$O$_{12}$ | 0.76 | 1.82 | 1.8 |
| LuGa$_5$O$_{12}$ | | 1.49 | |

*) Recalculated from $^{27}$Al NMR.

The Ga occupation numbers in the Lu$_3$Ga$_x$Al$_{5-x}$O$_{12}$ solid solutions can be directly determined from $^{71}$Ga NMR spectra. This can be done by measuring single crystal NMR spectra similar as in [29] because the $^{71}$Ga NMR spectrum from Ga tetrahedral sites in the Lu$_3$Ga$_x$Al$_{5-x}$O$_{12}$:Ce solid solutions is very broad (even the $^{71}$Ga central transition is spread over 300 kHz). The MAS technique cannot thus narrow it. The single crystal spectra taken at the frequency region of the central transition are shown in Fig. 3 at crystals orientation where spectral lines from the tetrahedral and octahedral Ga sites are most strongly separated.

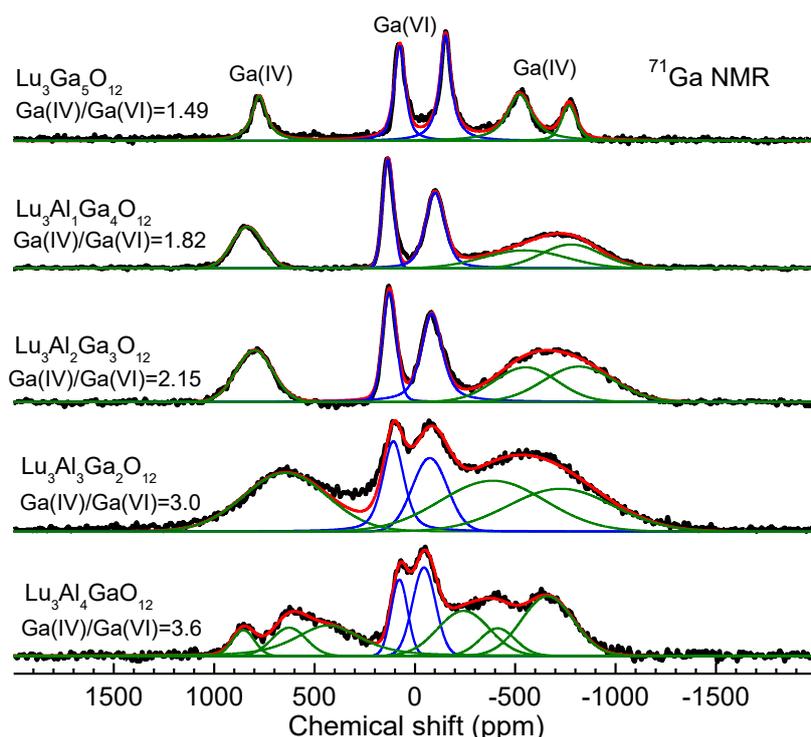

**Fig. 3**. $^{71}$Ga NMR spectra of Lu$_3$Ga$_x$Al$_{5-x}$O$_{12}$:Ce single crystals. The spectra are decomposed into separated lines from tetrahedral Ga(IV) (thin green lines) and octahedral Ga(VI) (thin blue lines) sites.

The Ga(IV)/Ga(VI) ratios calculated from the single crystals spectra are presented in Table 1 together with the Ga(IV)/Ga(VI) ratios recalculated from the Al(IV)/Al(VI) ratios assuming that the number of tetrahedral and octahedral sites in unit cell is 24 and 16, respectively.

To characterize more precisely the Al and Ga ions occupations, we have calculated fractional occupation numbers, defined as $f_{Al}^{tet} = Al(IV)/(Al(IV)+Al(VI))$ and $f_{Al}^{oct} = Al(VI)/(Al(IV)+Al(VI))$ for Al ions and similarly for Ga ions. These data are shown in Fig. 4. One can see in particular that at x = 1, only 20% of the Ga ions occupy the octahedral sites whereas 80% occupy the tetrahedral sites. Both $f_{Ga}^{tet}$ and $f_{Ga}^{oct}$ linearly tend to their normal values of 60 and 40% as the Ga concentration increases.

As it is evident from Fig. 4, the concentration ratio of the $Ga^{3+}$ ions in the tetrahedral and octahedral lattice sites linearly depends on the Ga content.

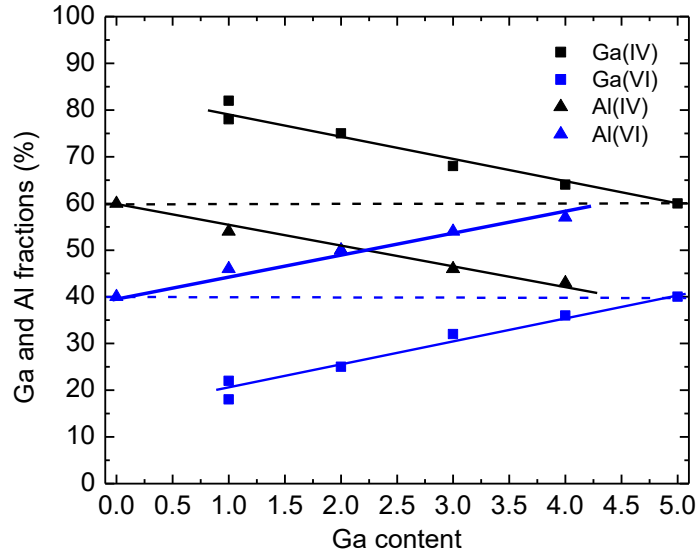

**Fig. 4.** Fractional occupation parameters of Ga and Al ions in the $Lu_3Ga_xAl_{5-x}O_{12}$:Ce mixed crystals as a function of the total Ga content. Dashed lines correspond to random distribution of Al and Ga over tetrahedral and octahedral sites, $f_{Al}^{tet} = f_{Ga}^{tet} = 60\%$ and $f_{Al}^{oct} = f_{Ga}^{oct} = 40\%$.

### *3.2. Photo- and thermally stimulated luminescence of $(Y,Lu)_3Ga_xAl_{5-x}O_{12}$*

The TSL glow curves measured for the $Y_1Lu_2(Ga,Al)_5O_{12}$:Ce and $Lu_3(Ga,Al)_5O_{12}$:Ce single crystals with different Ga contents (x) are shown in Figs. 5 and 6, respectively (see also Table 2). The corresponding trap depths $E_t$ determined by the partial heating method are indicated by filled circles. All the TSL glow curve peaks are complex and consist of several components. This is evident from the comparison of the TSL glow curves measured after irradiation at different temperatures $T_{irr}$ as well as from the difference in $E_t$ values in the range of each TSL peak. The total TSL intensity increases (by about an order of magnitude) with the increasing Ga content, reaches the maximum at x = 2 - 3 and then decreases. Most probably, the decrease is caused by the thermal quenching of the $Ce^{3+}$ emission whose temperature decreases with the increasing Ga content (Fig. 7). The intensity ratio of the low-temperature and high-temperature TSL peaks is also perturbed due to the temperature dependence of the $Ce^{3+}$ emission intensity. No $Ce^{3+}$ - related luminescence is observed in the samples with x = 5.

In Fig. 8, the dependences of the higher-temperature TSL peak position (filled triangles) and the corresponding trap depth value $E_t$ (empty circles) on the Ga content are presented. It is evident that both these dependences are similar to those shown in Fig. 1b for the epitaxial films of $(Lu,Gd)_3Ga_xAl_{5-x}O_{12}$:Ce and are close to linear. The same result is obtained also for the $Gd_3Ga_xAl_{5-x}O_{12}$:Ce single crystals in [33].

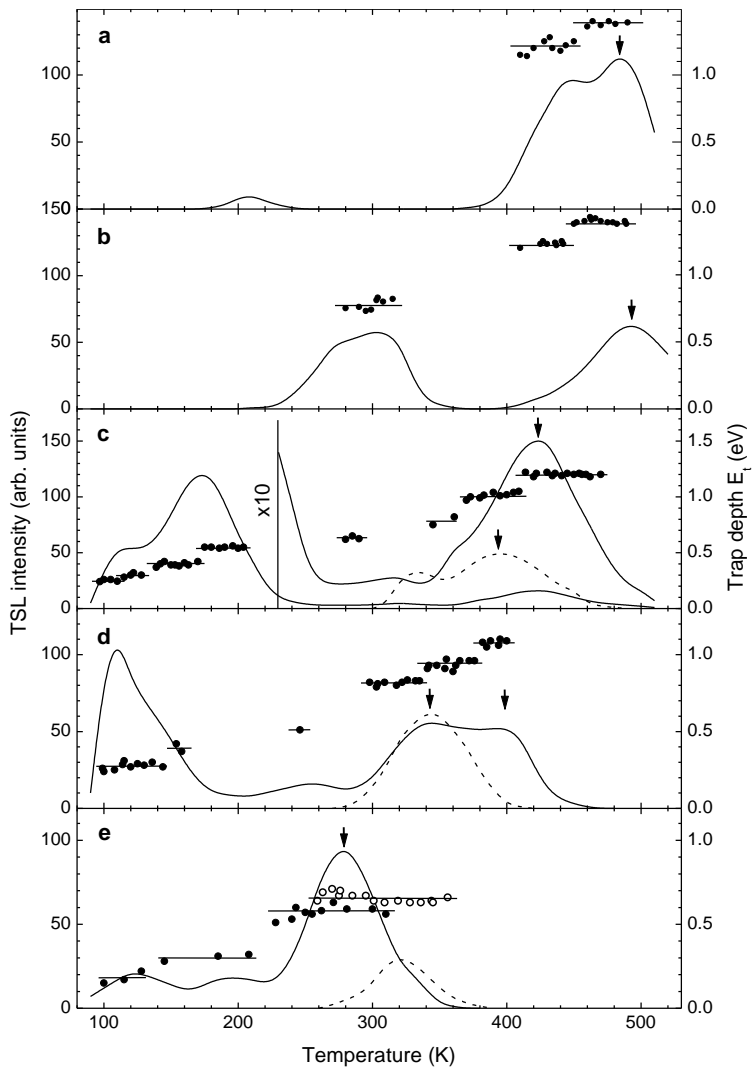

**Fig. 5.** TSL glow curves (lines) and trap depths $E_t$ (filled points) obtained for the single crystals of $Lu_2Y_1Ga_xAl_{5-x}O_{12}$:Ce with (a) x = 0, (b) x = 1, (c) x = 2, (d) x = 3, (e) x = 4 after irradiation at 85 K in the 4f - $5d_2$ absorption band of $Ce^{3+}$ ion (solid line) and at 295 K in the 4f - $5d_1$ absorption band of $Ce^{3+}$ ions (dashed line).

Similar dependence is obtained for the relative number of $Ga^{3+}$ ions substituting for $Al^{3+}$ ions in the tetrahedral lattice sites of the $Lu_3(Ga,Al)_5O_{12}$:Ce crystals determined by the NMR method (Fig. 4). As it follows from the data presented in Sec. 3.1, despite the larger ionic radius of $Ga^{3+}$ as compared with $Al^{3+}$, the $Ga^{3+}$ ions prefer to occupy first not the larger octahedral sites (as it has been believed before, see, e.g., [8,31]) but the smaller tetrahedral ones (see also [29,37] and references therein). However, as the Ga content increases, the number of $Ga^{3+}$ ions in the tetrahedral sites gradually decreases and the number of $Ga^{3+}$ ions in the octahedral sites increases resulting in the lowering of the CB bottom energy.

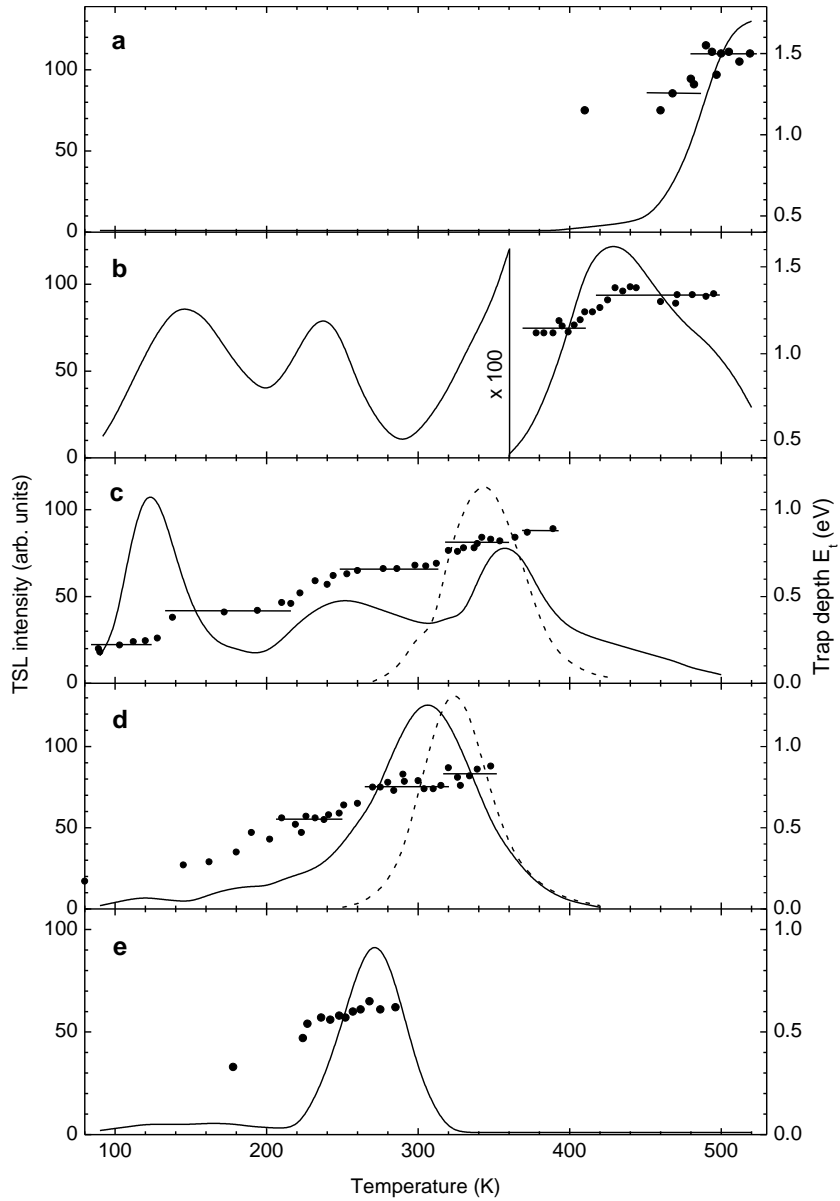

**Fig. 6.** TSL glow curves (lines) and trap depths $E_t$ (filled points) obtained for the single crystals of $Lu_3Ga_xAl_{5-x}O_{12}$:Ce with different Ga content: (a) x = 0, (b) x = 1, (c) x = 2, (d) x = 3, (e) x = 4 after

irradiation at 85 K in the 4f - $5d_2$ absorption band of $Ce^{3+}$ (solid line) and at 295 K in the 4f - $5d_1$ absorption band of $Ce^{3+}$ (dashed line).

**Table 2**. Activation energy $E_a$ of the TSL peaks creation under irradiation in the 4f – $5d_1$ absorption band of $Ce^{3+}$, activation energy $E_q$ of the luminescence thermal quenching, the TSL peaks position in $Lu_2Y_1(Ga,Al)_5O_{12}$:Ce and $Lu_3(Ga,Al)_5O_{12}$:Ce single crystals, and the energy difference $E_{dc}$ between the $5d_1$ level of $Ce^{3+}$ and the bottom of the CB estimated in this work and reported in [9] for $Lu_3(Ga,Al)_5O_{12}$:Ce with different Ga contents.

| Crystal | $E_a$, eV | $E_q$, eV | TSL peaks, K | $E_{dc}$, eV | $E_{dc}$, eV [9] |
|---|---|---|---|---|---|
| $Lu_2Y_1Ga_0Al_4O_{12}$:Ce | 0.97 | - | 207; 448; 484 | | |
| $Lu_2Y_1Ga_1Al_4O_{12}$:Ce | 0.87 | - | 275; 310; 494 | | |
| $Lu_2Y_1Ga_2Al_3O_{12}$:Ce | 0.33 | 0.15 | 118; 170; 365; 423 | | |
| $Lu_2Y_1Ga_3Al_2O_{12}$:Ce | 0.16 | 0.04; 0.30 | 110; 150; 342; 395 | | |
| $Lu_2Y_1Ga_4Al_1O_{12}$:Ce | 0.05 | 0.04; 0.16 | 122; 195; 280 | | |
| $Lu_3Ga_0Al_4O_{12}$:Ce | 0.91 | - | ~510 | 0.91 | 1.05 |
| $Lu_3Ga_1Al_4O_{12}$:Ce | 0.88 | - | 146; 238; 430 | 0.88 | 0.71 |
| $Lu_3Ga_2Al_3O_{12}$:Ce | 0.22 | 0.08; 0.23 | 124; 252; ~350 | 0.54 | 0.56 |
| $Lu_3Ga_3Al_2O_{12}$:Ce | 0.15 | 0.13; 0.34 | 120; 180; ~220; 306 | 0.30 | 0.25 |
| $Lu_3Ga_4Al_1O_{12}$:Ce | 0.04 | 0.03-0.06 | 272 | 0.04 | 0.09 |

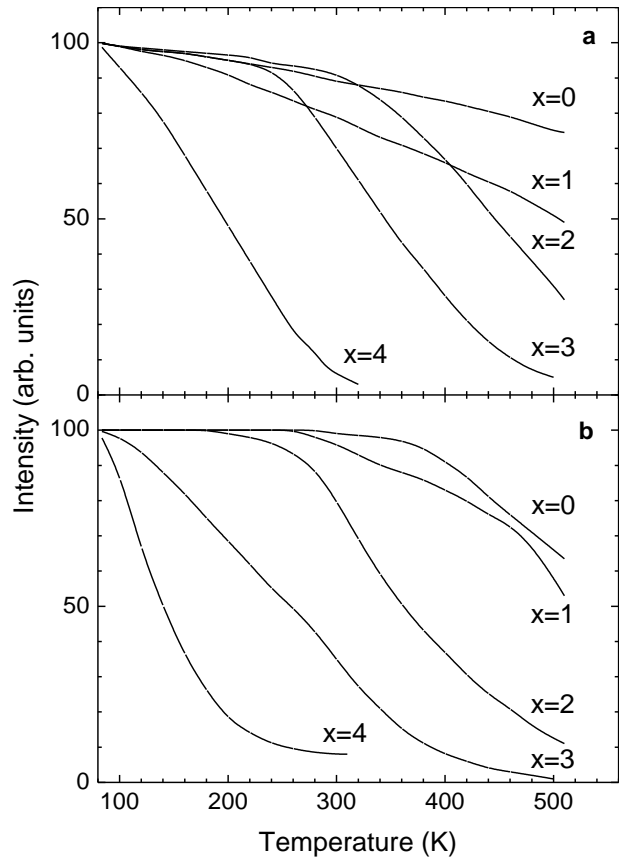

**Fig. 7.** Temperature dependences of the $Ce^{3+}$ emission intensity measured under excitation in the 4f - $5d_1$ absorption band of $Ce^{3+}$ ions for the single crystals of (a) $Lu_2Y_1Ga_xAl_{5-x}O_{12}$:Ce and (b) $Lu_3Ga_xAl_{5-x}O_{12}$:Ce with different Ga contents.

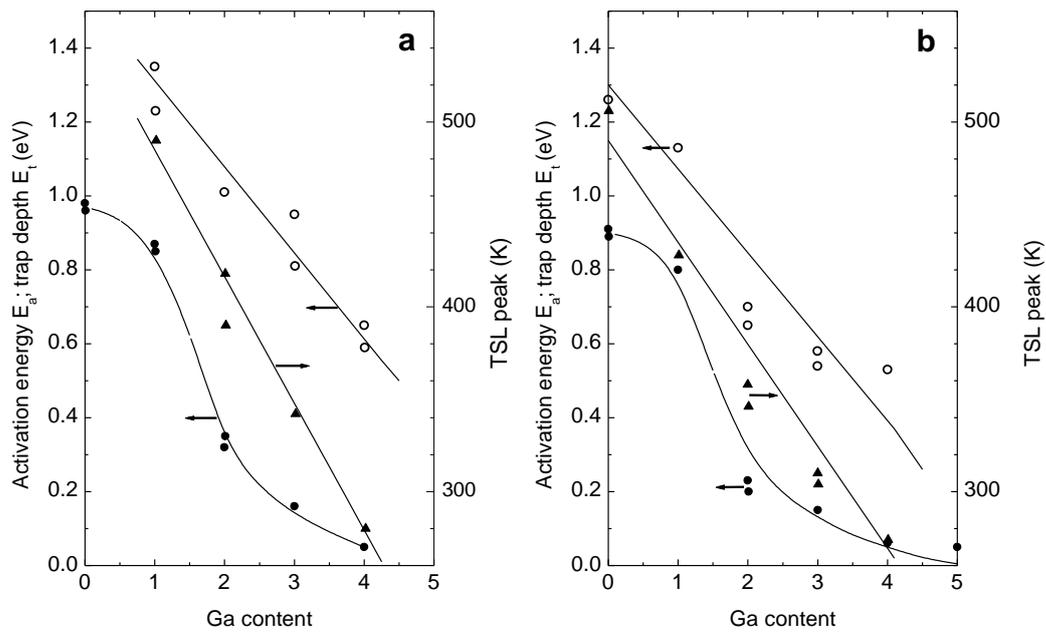

**Fig. 8.** Dependences of the activation energy $E_a$ of the TSL peaks creation (filled circles), the position of the high-temperature TSL peak (filled triangles), and the corresponding trap depth $E_t$ (empty circles) on the Ga content obtained for the single crystals of (a) $Lu_2Y_1Ga_xAl_{5-x}O_{12}$:Ce and (b) $Lu_3Ga_xAl_{5-x}O_{12}$:Ce.

Thus, the data presented in Figs. 4 and 8 clearly indicate that the dependence of the CB bottom energy on the Ga content is close to *linear*.

In Fig. 8, the dependences of the activation energy $E_a$ of the TSL peaks creation on the Ga content are also shown for the $Y_1Lu_2(Ga,Al)_5O_{12}$:Ce (a) and $Lu_3(Ga,Al)_5O_{12}$:Ce (b) single crystals (filled circles). For better comparison with the other multicomponent garnets, these data, as well as the data obtained for the $Gd_3Ga_xAl_{5-x}O_{12}$:Ce single crystals in [33], are all presented in Fig. 1a. As it is evident from Fig. 1a, the $E_a$ dependences on the Ga content practically coincide for all the multicomponent garnets studied up to now, and these dependences are *strongly nonlinear*. However, in the case the $E_a$ value be the energy distance $E_{dc}$ between the $5d_1$ excited level of $Ce^{3+}$ and the bottom of the CB, it should linearly decrease with the increasing Ga content in parallel to the $E_t$ dependence on the Ga content shown in Fig. 1b. In the samples without Ga and with a small Ga content, where the $E_a$ value is practically independent on the Ga content, the experimental $E_a$ value probably corresponds to the value of $E_{dc}$. If so, the $E_a$ value should change from about 0.9 eV at x = 1 to about 0.3 eV at x = 4. As it is evident from Fig. 1a, the experimental values of $E_a$ for the samples with x > 1 are located at much lower energies.

Taking these data into account, we have to suggest that the activation energy $E_a$ for the TSL peak creation under the 4f - $5d_1$ excitation is not always equal to the energy distance $E_{dc}$ between the excited $5d_1$ level of $Ce^{3+}$ and the bottom of the CB. The observed peculiarities of the luminescence thermal quenching and photostimulated defects creation processes can be explained by the presence in the investigated crystals with sufficiently large Ga content of additional energy levels located between the $5d_1$ excited level of $Ce^{3+}$ and the bottom of the CB. In this case, thermally stimulated electron transitions from the excited $5d_1$ level of $Ce^{3+}$, resulting in the defects creation, can take place not only to the CB but also to these additional defect levels, and $E_a$ can be considered as the energy difference between the $5d_1$ level and the defect level (d). In Fig. 9, the positions of the $5d_1$ and $5d_2$ excited levels of $Ce^{3+}$ with respect to the 4f ground state level and the position of the defect level d with respect to the $5d_1$ level of $Ce^{3+}$ are schematically presented for the $Lu_3(Ga,Al)_5O_{12}$:Ce crystals. The approximate position of the CB bottom is shown by a dashed line taking into account the experimental data indicating that (i) in the crystals with x = 0 and x = 1, the $E_a$ value (~ 0.9 eV, see Table 2) corresponds most probably to the energy distance $E_{dc}$ between the $5d_1$ level of $Ce^{3+}$ and the bottom of the CB, i.e, the $E_{dc}$ value cannot be smaller than ~0.9 eV; (ii) in the crystal with x = 2, the $5d_2$ level is located inside the CB, i.e., the $E_{dc}$ value cannot be smaller than 0.56 eV; (iii) in the crystal with x=4, the $E_{dc}$ value cannot be smaller than $E_a$= 0.04 eV; (iv) the CB bottom energy decreases linearly with the increasing Ga content. The estimated $E_{dc}$ values are shown in Table 2. For comparison, the values of $E_{dc}$ determined in [9] are also presented. As it is evident from Table 2, for the crystals with x=2 and x=3, the $E_{dc}$ and $E_a$ values are strongly different. The $E_{dc}$ values estimated in this work are rather close to those reported in [9] for $Lu_3(Ga,Al)_5O_{12}$:Ce.

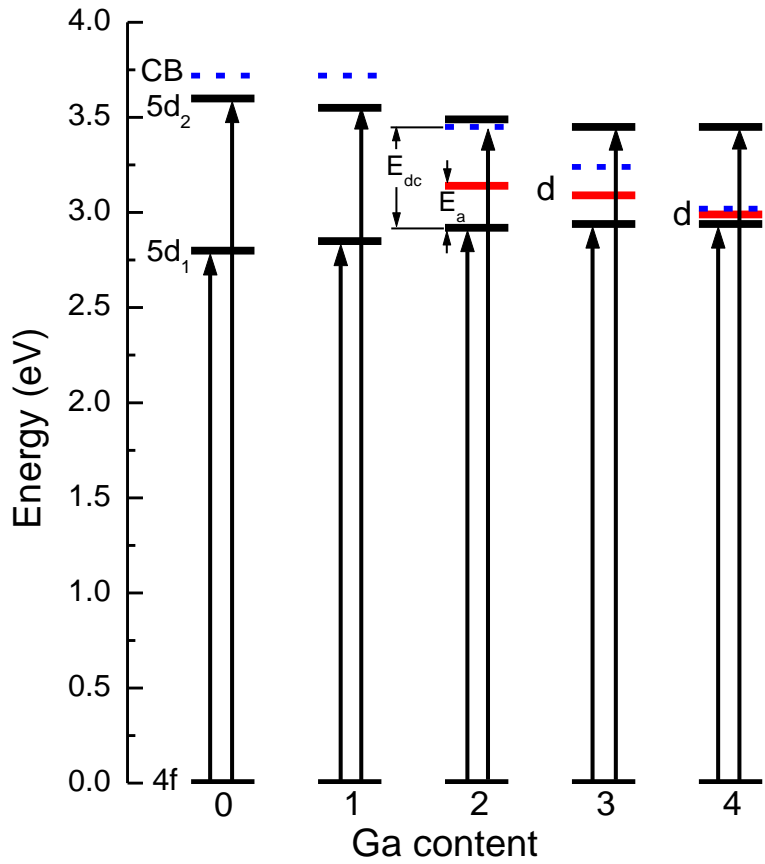

**Fig. 9.** Positions of the $5d_1$ and $5d_2$ excited levels of $Ce^{3+}$ and the defect level d (solid lines) and the suggested position of the CB bottom (dashed line) with respect to the 4f ground state level of $Ce^{3+}$ in $Lu_3Ga_xAl_{5-x}O_{12}$:Ce single crystals with different Ga contents. The 4f - $5d_1$ and 4f - $5d_2$ transitions of $Ce^{3+}$ ions are shown by arrows. $E_a$ is the energy difference between the $5d_1$ and d levels. $E_{dc}$ is the estimated energy difference between the $5d_1$ level of $Ce^{3+}$ and the CB bottom.

Some other experimental data can also confirm this suggestion. For example, in case the electrons can be thermally released only into the CB from both the $5d_2$ and the $5d_1$ excited state of $Ce^{3+}$, the TSL glow curves should be independent of the excitation energy. However, in the crystals with large enough Ga content, the high-temperature TSL peaks observed after $5d_2$ irradiation (when an electron is surely released into the CB) and after $5d_1$ excitation at the appropriate temperature are sometimes different (see, e.g., Figs. 5d, 5c). Indeed, the TSL peak at about 400 K observed in the $Y_1Lu_2Ga_3Al_2O_{12}$:Ce crystal after $5d_2$ excitation does not appear after $5d_1$ excitation (Fig. 5d).

In case the electrons from the 5d excited states of $Ce^{3+}$ could be thermally released only into the CB, the activation energies for the luminescence thermal quenching ($E_q$) and the TSL peaks creation ($E_a$) should coincide. However, like in the samples studied in [25,30,33], the $E_a$ values are sometimes strongly different from the $E_q$ values determined from temperature dependence of the emission intensity shown in Fig. 7 (see Table 2). Some examples of the lnI - 1/T dependences, from where the $E_q$ values were determined, are shown in Fig. 10.

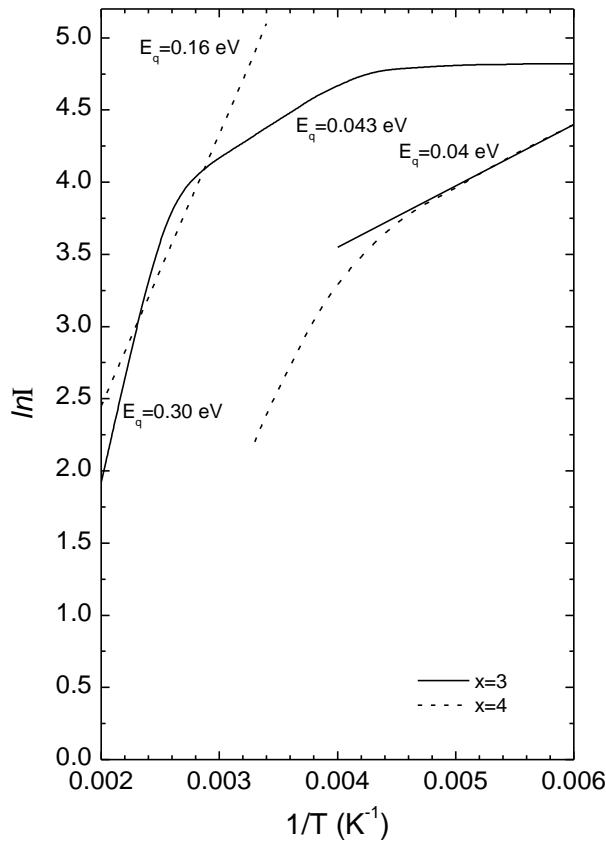

**Fig. 10.** The lnI - 1/T dependences obtained for the $Lu_2Y_1Ga_3Al_2O_{12}$:Ce (solid line) and $Lu_2Y_1Ga_4Al_1O_{12}$:Ce (dashed lines) single crystals and used for calculation of the activation energies $E_q$.

Let us consider now a possible origin of the above-mentioned defect level. The concentration of the main intrinsic crystal lattice defects (non-compensated vacancies and antisites) in the epitaxial films is always much smaller as compared with the single crystals (see, e.g., [20] and references therein). Despite of that, the $E_a$ dependences on the Ga content coincide in the epitaxial films and single crystals (Fig. 1a). Besides, the TSL investigations carried out in [30] have shown that in the single crystals of $Gd_3(Ga,Al)_5O_{12}$:Ce and $Gd_3(Ga,Al)_5O_{12}$:Ce,Mg of the same composition, the concentrations of intrinsic defects are strongly different. Indeed, in the $Gd_3(Ga,Al)_5O_{12}$:Ce,Mg single crystal, the TSL intensity is by about two orders of magnitude weaker as compared with that in the $Gd_3(Ga,Al)_5O_{12}$:Ce single crystal and is similar to that in the $Gd_3(Ga,Al)_5O_{12}$:Ce epitaxial film of the same composition. However, the $E_a$ values obtained for all the three samples practically coincide. As the dependences of the $E_a$ values on the Ga content are very close in the samples containing strongly different concentrations of intrinsic defects, the above-mentioned defect level cannot be connected with the intrinsic defects.

Therefore, we suggest that the above-mentioned defect level might arise from the $Ga^{3+}$ ions substituting for $Al^{3+}$ ions and perturbed by the nearest neighboring $Ce^{3+}$ ions. Then thermally stimulated electron transitions from the excited $5d_1$ level of $Ce^{3+}$ can occur not to the CB but to the $Ga^{3+}$ - related level, and the activation energy $E_a$ corresponds to the energy distance between the $5d_1$ excited level of $Ce^{3+}$ and the $Ga^{3+}$ - related level. According to [4,38], $Ga^{3+}$ energy levels are placed around the very bottom of the CB, so that their participation in the creation of electron traps can be expected. Besides, a bend in the CB in the neighborhood of a defect can occur. The possibility of this effect has been theoretically considered in [39] basing on the Kronig-Penney model.

The trapping of the electrons, thermally released from the excited $Ce^{3+}$ ions, by the perturbed $Ga^{3+}$ ions can result in the formation of electron $Ga^{2+}$ - centers perturbed by $Ce^{3+}$. The efficiency of the electron trapping could be different for the $Ga^{3+}$ ions located in different (tetrahedral or octahedral) crystal lattice sites. Thermally stimulated destruction of the electron $Ga^{2+}$ centers should result in the appearance of the TSL peaks. As it is evident from [25], the intensity ratio of the low-temperature and high-temperature TSL peaks strongly depends on the Ga content. In the samples with $x = 0$ and $x = 0.81$, no low-temperature TSL peaks are observed after irradiation at 85 K, but at $x = 1.41$, the intense peak around 150 K appears. As the Ga content increases, this peak is shifting to lower temperatures and becomes the dominating one in the crystals with $x = 2$. The same result is also obtained in the single crystals of $Y_1Lu_2Ga_xAl_{5-x}O_{12}$:Ce, $Lu_3Ga_xAl_{5-x}O_{12}$:Ce (see Figs. 5 and 6) and $Gd_3Ga_xAl_{5-x}O_{12}$:Ce [33]. Probably, this complex TSL peak arises just from the electron $Ga^{2+}$ - related centers.

From the comparison of the TSL curves, measured after 4f - 5d$_2$ irradiation at 85 K with different irradiation durations t$_{irr}$, it is evident that the traps, responsible for the lower-temperature TSL peaks, become filled up to the saturation much more quickly as compared with the traps responsible for the high-temperature peaks (Fig. 11, see also Fig. 12). This dose dependence is characteristic for the closely located defects.

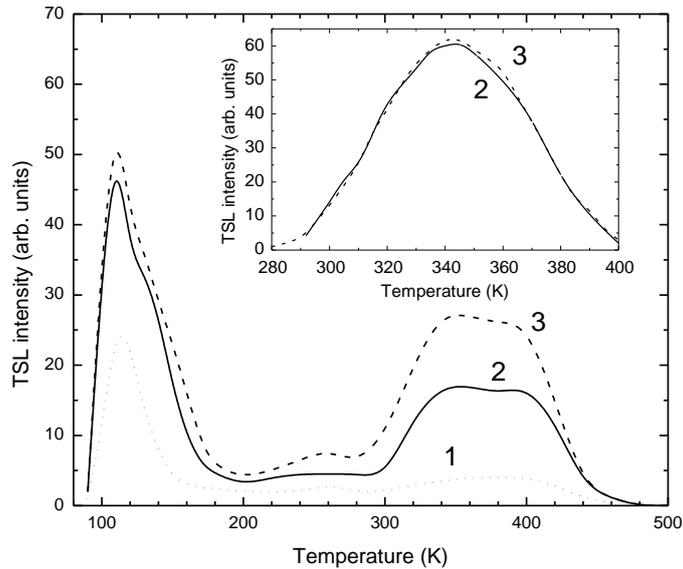

**Fig. 11.** TSL glow curves measured for the Lu$_2$Y$_1$Ga$_3$Al$_2$O$_{12}$:Ce crystal after irradiation at 85 K in the 4f - 5d$_2$ absorption band of Ce$^{3+}$ for 1 min (curve 1), 5 min (curve 2), and 10 min (curve 3). In the inset, the TSL glow curves measured after irradiation at 295 K in the 4f-5d$_1$ absorption band of Ce$^{3+}$ for 5 min (curve 2), and 10 min (curve 3).

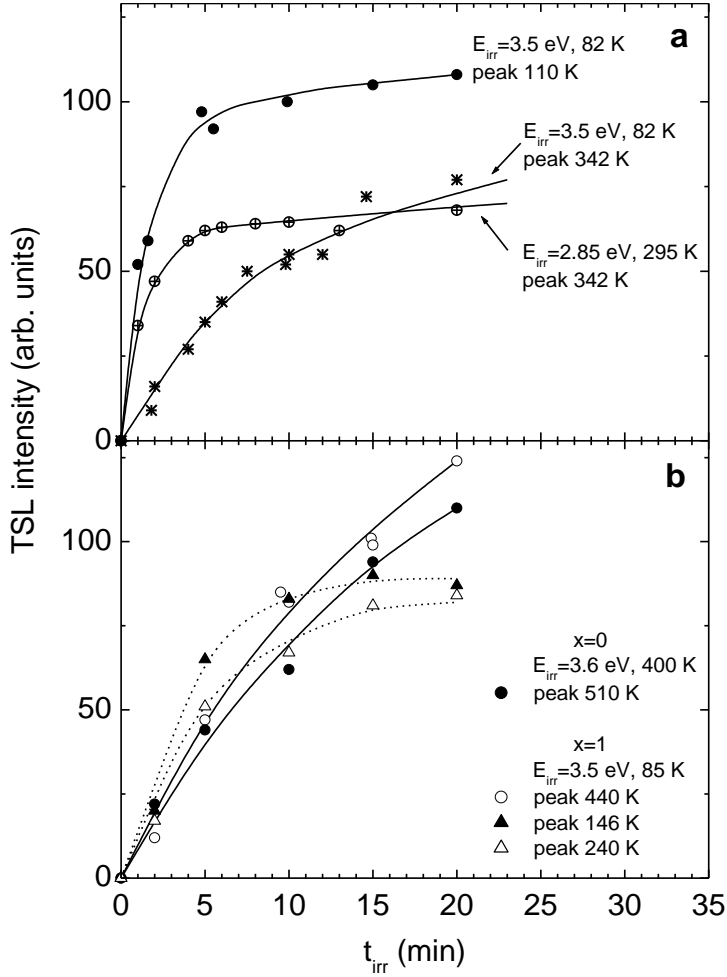

**Fig. 12.** Dependences of the TSL intensity on the irradiation duration $t_{irr}$ measured for the single crystals of (a) $Lu_2Y_1Ga_3Al_2O_{12}$:Ce and (b) $Lu_3Ga_xAl_{5-x}O_{12}$:Ce. The selected TSL peaks and irradiation conditions are shown in the legends.

The stable $Ga^{2+}$ centers with visible photoluminescence have been intensively studied in alkali halides (see, e.g., [40,41]). However, our attempts to find the luminescence of $Ga^{2+}$ centers in the investigated crystals were unsuccessful. This can be caused by the presence of intense $Ce^{3+}$ - related absorption and emission bands in the energy range characteristic for $Ga^{2+}$ centers. Besides, if the ground state of $Ga^{2+}$ centers is located well below the CB, then their excited levels should be located inside the CB. In this case, no intra-centre luminescence of $Ga^{2+}$ in the investigated crystals could be observed. However, electron $Ga^{2+}$ centers are paramagnetic as $Ga^{2+}$ ion has $4S^1$ outer electron shell with the electron spin 1/2. Therefore, we have tried to detect these centers by EPR in the $Lu_3(Ga,Al)_5O_{12}$:Ce crystals with different Ga contents irradiated in the absorption bands of $Ce^{3+}$ at the temperatures where the electrons can be optically released from $Ce^{3+}$ ions and subsequently trapped at the $Ga^{3+}$ - related traps. However, out attempts to find EPR signal from $Ga^{2+}$ ions was unsuccessful. It does not unambiguously mean that such ions can not exist in the crystals. Simply, the $Ga^{2+}$ EPR signals could be weak due to broad spectral lines and/or large hyperfine splitting as Ga has two isotopes with non-zero nuclear magnetic moment: $^{69}$Ga (I=3/2, natural abundance 60.4%) and $^{71}$Ga (I=3/2, natural abundance 39.6%). It is also expected a strong superhyperfine interaction of electron spin with nuclear spins of surrouding Al and Lu ions which possess large nuclear magnetic moments. For instance, in NaCl [42] the hyperfine constant for $Ga^{2+}$ impurity ion is 19-24 GHz and the width of superhyperfine structure is more than 100 G. All these factors make EPR identification of $Ga^{2+}$ ions quite difficult task.

However, under irradiation of the investigated crystals in the $Ce^{3+}$ - related absorption bands with laser at 442 nm (2.8 eV), the transformation of $Fe^{3+}$ ions into $Fe^{2+}$ ions has been observed by EPR. $Fe^{3+}$ ions as a background impurity always exist in garnet crystals and can act as electron traps as well. Thus, in principle, $Fe^{3+}$ ions can also be responsible for the appearance of the above-mentioned defect level located below the CB. According to [43], in Ga-rich garnet crystals, $Fe^{3+}$ predominantly locates at the octhahedral sites. As an example, Fig. 13 shows change of the $Fe^{3+}$ EPR spectrum after irradiation at 442 nm at room temperature for three crystals with the Ga contents x = 5, 4 and 2.5. One can see that $Fe^{3+}$ spectral lines marked in the figure by arrows substantially change their intensity after irradiation of the samples with x = 5 and 4. However, this effect is negligibly small in the sample with x = 2.5 and lower Ga contents. After the laser light is switched off, the $Fe^{3+}$ EPR signal slowly recovers to its initial intensity.

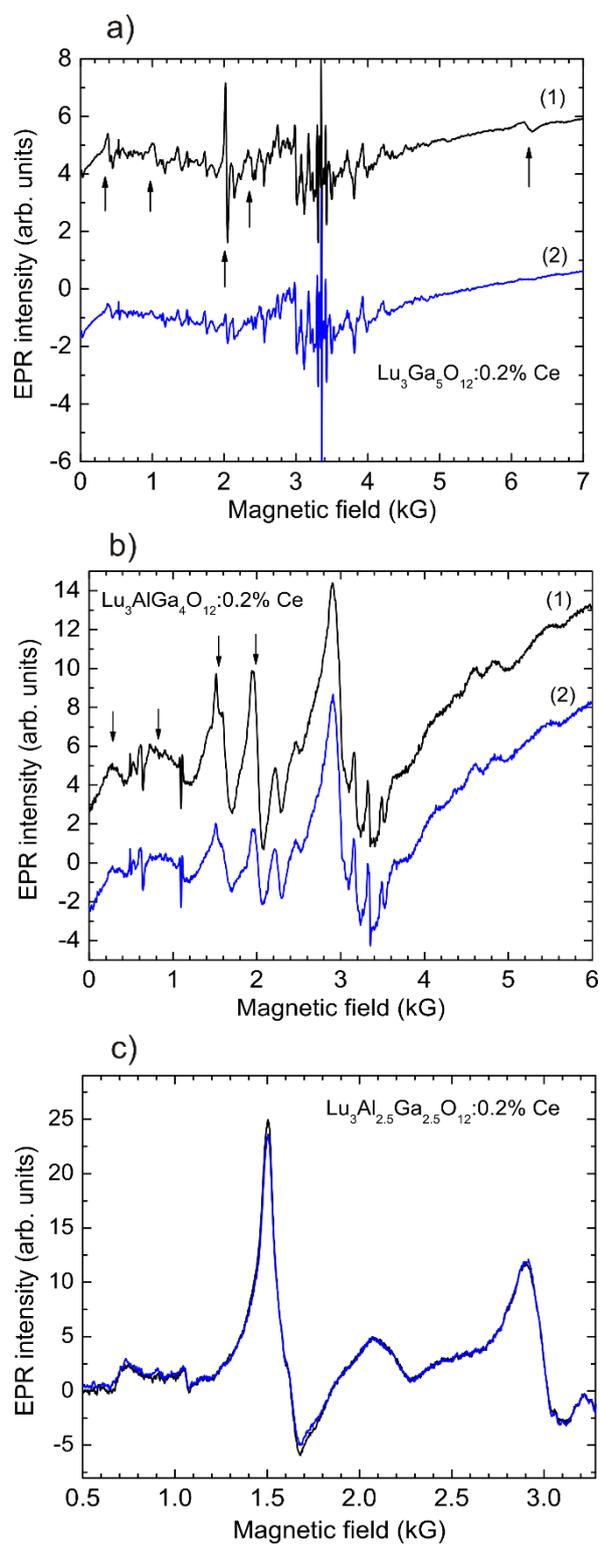

**Fig. 13**. EPR spectra measured in three garnet crystals a) $Lu_3Ga_5O_{12}$, b) $Lu_3AlGa_4O_{12}$ and c) $Lu_3Al_{2.5}Ga_{2.5}O_{12}$ doped by 0.2% Ce before (1) and after (2) 442 nm irradiation at 297 K for 15 min. The spectra are measured at 297 K. Arrows show spectral lines attributed to $Fe^{3+}$. Only a weak change in the $Fe^{3+}$ intensity is detected in the $Lu_3Ga_{2.5}Al_{2.5}O_{12}$ crystal (spectra before and after irradiation practically coinside, see panel c)).

We further performed measurements of temperature and time characteristics of the observed effect. Fig. 14 shows the $Fe^{3+}$ EPR intensity decay and rise at selected temperatures for the sample with x = 4. The decay and rise curves were fitted by two-exponent functions as it is illustrated by solid red line for 297 K. At this temperature, the decay times are $\tau_1$ = 7.2 s, $\tau_2$ = 48 s and the rise times $\tau_1$ = 71 s, $\tau_2$ = 647 s. These times only weakly change with temperature down to 100 K suggesting the presence of a weak thermally activated electron excitation process. Similar measurements were done for $Lu_3Ga_5O_{12}$:0.2% Ce but not in the same detail as for the former sample. Much stronger change as compared to decay and rise times is seen in the decrease of $Fe^{3+}$ concentration under irradiation at different temperatures. The largest decrease in $Fe^{3+}$ concentration takes place at 297 K and it gradually decreases with the temperature lowering.

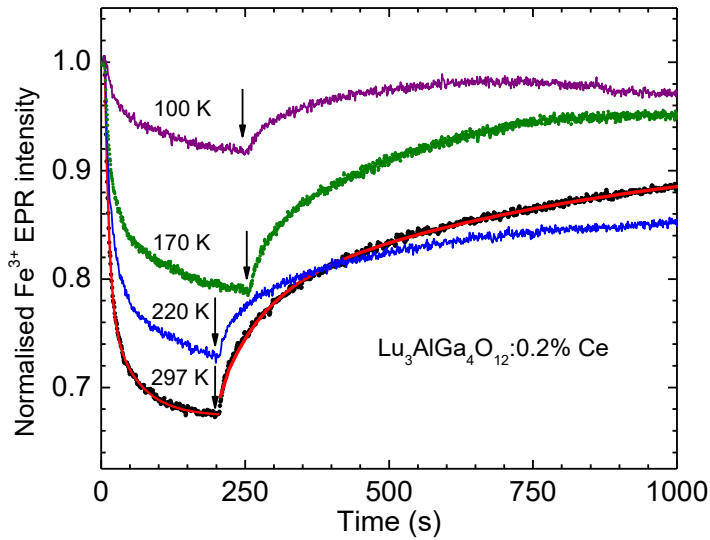

**Fig. 14**. Normalized $Fe^{3+}$ intensity as a function of time when the laser light is switched on at t = 0 and switched off (shown by arrows) at selected temperatures in $Lu_3AlGa_4O_{12}$:0.2% Ce. The solid red line for the data at 297 K is theoretical fit.

The observed phenomenon can be undersood assumingt that under 442 nm excitation of $Ce^{3+}$ ions, thermally stimulated electron transitions take place from the excited $5d_1$ level of $Ce^{3+}$ not only to the $Ga^{3+}$- but also to the $Fe^{3+/2+}$ - related levels resulting in formation of $Fe^{2+}$ states which are EPR-silent in garnet crystals. Their EPR spectrum can not be visible in the X-band used in our measurements. However, the possibility of the $Fe^{3+} + e \rightarrow Fe^{2+}$ process was reported in many papers dealing with optical absorption spectra in YAG, YGAG and YIG crystals (see, e.g. [44-46]). In all these publications, the electron transfer from the oxygen 2p band to $Fe^{3+}$, taking place with the energy of 4.6 - 4.8 eV, was considered. We can thus attribute the decrease of the $Fe^{3+}$ concentration with recharge of these ions to $Fe^{2+}$. It should be noted that, unlike the previous works, the $Fe^{3+}$ - $Fe^{2+}$ transformation is observed in this paper under irradiation in the absorption band of $Ce^{3+}$ ions, and it takes place due to thermally stimulated (probably, tunnelling) electron transition from the excited $5d_1$ level of $Ce^{3+}$ to the excited state of $Fe^{3+}$.

Obviously, the electron transfer from the excited $Ce^{3+}$ $5d_1$ level needs some thermal activation energy $E_a$. It can be determined from the $Fe^{3+}$ concentration decay times as a function of temperature. Assuming Arrhenius law for the electron thermal excitation rate, the experimentally determined decay times can be fitted by the following expression:

$$\tau^{-1}(T) = \tau_0^{-1} \exp(-E_a/kT) + \tau_r^{-1} \tag{1}$$

Here $\tau_0^{-1}$ is the frequency factor of electron escape and $\tau_r^{-1}$ takes into account temperature independent process. Fig. 15 shows the fit of experimental data to Eq. (1) from which $E_a$ = 0.064 eV was determined for faster decay component in $Lu_3AlGa_4O_{12}$ sample. Similar fit for the slower decay component gives $E_a$ = 0.043 eV. These $E_a$ values, taking into account large standard error of 0.03 eV, are comparable with the activation energy obtained for the same sample from the TSL data ($E_a$ = 0.042 eV, see Fig. 8b). Note that the presence of two thermally activated processes may be attributed to two different space configurations of $Ce^{3+/4+}$ – $Fe^{3+/2+}$ pairs or different distances between partners in the pair. In the sample with x = 2.5, the change in the $Fe^{3+}$ concentration at room temperature is negligible (Fig. 13, panel c) indicating a larger value of $E_a$.

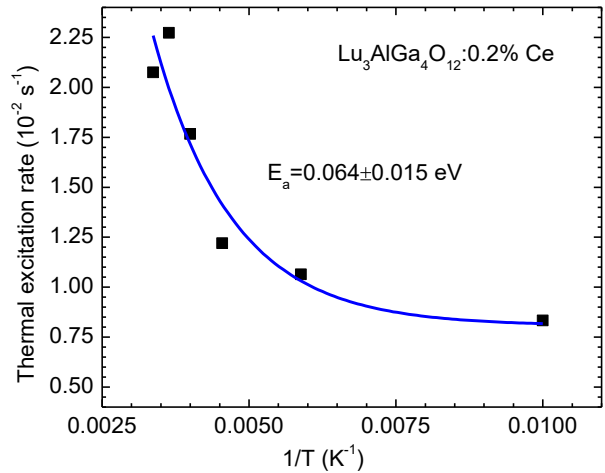

**Fig, 15**. Plot of electron thermal excitation rate determined from $Fe^{3+}$ concentration decay under 442 nm irradiation of $Lu_3AlGa_4O_{12}$:0.2% Ce crystal as a function of reciprocal temperature.

In order to confirm that the electron transfer from $Ce^{3+}$ to $Fe^{3+}$ takes place without a participiation of the conduction band we also irradiated the crystals with photon energy larger than the distance from Ce 4f levels to the CB at different temperatures. No change in $Fe^{3+}$ concentration was found as well as even after X-ray irradiation at 77 K indicating that most probably other impurity ions or defects are responsible for the observed TSL peaks.

Thus, the EPR data confirm that thermally stimulated electron transfer from the excited $5d_1$ level of $Ce^{3+}$ to other ions is really possible and takes place with the activation energy depending on the type of the ion. However, the huge difference in the thermal stablity of the traps responsible for TSL peaks (see Fig. 1b) and thermal stability of $Fe^{2+}$ ion, as well as a strong dependence of the $E_a$ value on the Ga content in all the investigated multicomponent garnets (Fig. 1a) allow us to suggest that the defect level located between the $5d_1$ level of $Ce^{3+}$ and the CB is most probably connected with Ga and arises from $Ga^{3+}$ ions perturbed by close $Ce^{3+}$ ions.

## 4. Conclusions

The increasing Ga content is found to result in the decreasing $E_a$ value, increasing TSL intensity (observed at least up to x=3), decreasing temperature of the luminescence quenching, low-temperature shift of the TSL peaks and decrease of the corresponding trap depths. It results also in the increasing concentration of $Ga^{3+}$ ions in the octahedral $Al^{3+}$ sites as compared to the tetrahedral ones, resulting in the lowering of the CB bottom energy. The dependences of the TSL peaks positions, corresponding trap depths, and fractional occupation parameters of Ga and Al ions on the Ga content are close to *linear*. This fact indicates that the bottom of the CB is shifting linearly with the increasing Ga content.

However, in both the epitaxial films and the single crystals of different composition and origin, having strongly different concentrations of non-compensated vacancies and antisite defects, the $E_a$ dependences on the Ga content practically coincide and are strongly *nonlinear*. These data allow us to suggest that in the $Ga^{3+}$ - containing multicomponent garnets with sufficiently large Ga content, the activation energy $E_a$ is not equal to the energy distance between the $5d_1$ excited level of $Ce^{3+}$ and the bottom of the CB. We suggest that $E_a$ is the energy distance between the $5d_1$ level of $Ce^{3+}$ and a defect level located between the $5d_1$ level and the CB. This defect level is suggested to arise from the $Ga^{3+}$ ions perturbed by the nearest neighboring $Ce^{3+}$ ions.

The thermally stimulated release of electrons from the excited $Ce^{3+}$ ions and their subsequent trapping at the perturbed $Ga^{3+}$ ions can result in the appearance of the perturbed electron $Ga^{2+}$ centers. These centers can be responsible for the TSL peaks located in the 100 - 150 K temperature range.

Thus, the presented data indicate much more complex structure of the bands and the defect levels in the band gap as compared with that suggested before as well as a strong influence of $Ga^{3+}$ ions on this structure. The peculiarities of the $Ga^{3+}$ - containing multicomponent garnets are caused by the fact that, unlike the rare-earth $Y^{3+}$, $Lu^{3+}$, $Gd^{3+}$ ions and $Al^{3+}$, the perturbed $Ga^{3+}$ ions can act as effective traps for electrons. These data also indicate that in the multicomponent garnets, Ga can play not only a positive but also a negative role (the appearance of additional defects, reduced activation energy for the photostimulated defects creation and luminescence thermal quenching) which increases with the increasing Ga content.

**Acknowledgments**

The work was supported by the Institutional Research Funding IUT02-26 of the Estonian Ministry of Education and Research and the project 17-09933S of the Czech Science Foundation.